\documentclass[journal,twoside]{IEEEtran}

\usepackage{verbatim}
\usepackage[T1]{fontenc}
\usepackage{amsmath}
\usepackage[cmintegrals]{newtxmath}
\usepackage{bm}
\usepackage{nicefrac}
\usepackage[utf8]{inputenc}
\usepackage{todonotes}
\usepackage{graphicx}
\graphicspath{{./leaks/}}
\DeclareGraphicsExtensions{.pdf,.eps,.png,.jpeg}
\setkeys{Gin}{width=\linewidth}
\newcommand{\berkertitle}{Arraymetrics: Authentication Through Chaotic~Antenna Array Geometries}
\usepackage{color} 
\definecolor{mygreen}{RGB}{28,172,0} 
\definecolor{mylilas}{RGB}{170,55,241}

\usepackage{breqn}
\setkeys{breqn}{labelprefix={eq:}}

\newcommand{\norm}[2]{\left\lVert#1\right\rVert_{#2}}

\newcommand*{\matr}[1]{\bm{#1}}

\newcommand*{\herm}[1]{{#1}^{\mkern-1.5mu\mathsf{H}}}

\usepackage{dblfloatfix}
\usepackage{balance}
\usepackage{cite}
\usepackage{jabbrv}
\usepackage{algpseudocode}

\newcounter{tempEquationCounter} 
\newcounter{thisEquationNumber}

\usepackage[acronym]{glossaries}
\makeglossaries

\newacronym{NSN}{NSN}{narrow subcarrier numerology}
\newacronym{WSN}{WSN}{wide subcarrier numerology}
\newacronym{FMCW}{FMCW}{frequency modulated continuous wave}
\newacronym{fdm}{FDM}{frequency division multiplexing}
\newacronym{THz}{THz}{terahertz}
\newacronym{OFDM}{OFDM}{orthogonal \gls{fdm}}
\newacronym{dft}{DFT}{discrete Fourier transform}
\newacronym{dftsofdm}{DFT-s-OFDM}{\gls{dft} spread \gls{OFDM}}
\newacronym{wOFDM}{W-OFDM}{windowed-\gls{OFDM}}
\newacronym{fnd}{FND}{fractional numerology domain}
\newacronym{OFDMA}{OFDMA}{orthogonal frequency division multiple access}
\newacronym{eMBB}{eMBB}{enhanced mobile broadband}
\newacronym{uRLLC}{uRLLC}{ultra-reliable low-latency communication}
\newacronym{MBRLLC}{MBRLLC}{mobile broadband reliable low latency communication}
\newacronym{HCS}{HCS}{human-centric services}
\newacronym{muRLLC}{muRLLC}{massive \gls{uRLLC}}
\newacronym{lan}{LAN}{local area network}
\newacronym{wlan}{WLAN}{wireless \gls{lan}}
\newacronym{rc}{RC}{raised cosine}
\newacronym{osi}{OSI}{Open System Interconnection}
\newacronym{rrc}{RRC}{root \gls{rc}}
\newacronym{ini}{INI}{inter-numerology interference}
\newacronym{scs}{$\Delta{f}$}{subcarrier spacing}
\newacronym{pifa}{PIFA}{planar inverted-F antenna}
\newacronym{cdf}{CDF}{cumulative distribution function}
\newacronym{bt}{BT}{bandwidth-time}
\newacronym{CoMP}{CoMP}{coordinated multipoint}
\newacronym{poa}{PoA}{price of anarchy}
\newacronym{LTE}{LTE}{Long Term Evolution}
\newacronym{LTE-A}{LTE-A}{LTE-Advanced}
\newacronym{3GPP}{3GPP}{$3^{rd}$ Generation Partnership Project}
\newacronym{FFR}{FFR}{fractional frequency reuse}
\newacronym{SFR}{SFR}{soft frequency reuse}
\newacronym{CCI}{CCI}{co-channel interference}
\newacronym{ICIC}{ICIC}{inter-cell interference coordination}
\newacronym{eICIC}{eICIC}{enhanced \gls{ICIC}}
\newacronym{TP}{TP}{transmission point}
\newacronym{CSI}{CSI}{channel state information}
\newacronym{eNB}{eNB}{eNodeB}
\newacronym{RAN}{RAN}{radio access network}
\newacronym{C-RAN}{C-RAN}{cloud-RAN}
\newacronym{F-RAN}{F-RAN}{fog-RAN}
\newacronym{RRH}{RRH}{remote radio head}
\newacronym{BBU}{BBU}{baseband unit}
\newacronym{CS}{CS}{coordinated scheduling}
\newacronym{CB}{CB}{coordinated beamforming}
\newacronym{JT}{JT}{joint transmission}
\newacronym{DPS}{DPS}{dynamic point selection}
\newacronym{TTI}{TTI}{transmission time interval}
\newacronym{ue}{UE}{user equipment}
\newacronym{RNTP}{RNTP}{relative narrowband transmitted power}
\newacronym{OI}{OI}{overload indicator}
\newacronym{HII}{HII}{high interference indicator}
\newacronym{HG}{HG}{Hermite-Gaussian}
\newacronym{PMI}{PMI}{precoding matrix indicator}
\newacronym{RI}{RI}{rank indicator}
\newacronym{hetnet}{HetNet}{heterogeneous network}
\newacronym{mmWave}{mmWave}{millimeter wave}
\newacronym{sinr}{SINR}{signal-to-interference-plus-noise ratio}
\newacronym{sigint}{SIGINT}{signals intelligence}
\newacronym{awgn}{AWGN}{additive white Gaussian noise}
\newacronym{ml}{ML}{machine learning}
\newacronym{SC}{SC}{single-carrier}
\newacronym{GB}{GB}{guard band}
\newacronym{GD}{GD}{guard duration}
\newacronym{DL}{DL}{downlink}
\newacronym{fa}{FA}{false alarm}
\newacronym{FFT}{FFT}{fast Fourier transform}
\newacronym{IFFT}{IFFT}{inverse fast Fourier transform}
\newacronym{CP}{CP}{cyclic prefix}
\newacronym{GBR}{GBR}{guaranteed bit-rate}
\newacronym{gfdm}{GFDM}{generalized \gls{OFDM}}
\newacronym{RSRP}{RSRP}{reference signal received power}
\newacronym{RSRQ}{RSRQ}{reference signal received quality}
\newacronym{CSI-RSRP}{CSI-RSRP}{CSI-reference signal received power}
\newacronym{se}{SE}{spectral efficiency}
\newacronym{2G}{2G}{second-generation}
\newacronym{3G}{3G}{third-generation}
\newacronym{4G}{4G}{fourth generation}
\newacronym{5g}{5G}{fifth generation}
\newacronym{6G}{6G}{sixth generation}
\newacronym{papr}{PAPR}{peak-to-average power ratio}
\newacronym{pa}{PA}{power amplifier}
\newacronym{pe}{PE}{power efficiency}
\newacronym{wdm}{WDM}{waveform division multiplexing}
\newacronym{noma}{NOMA}{non-orthogonal multiple access}
\newacronym{mimo}{MIMO}{multiple-input multiple-output}
\newacronym{mamimo}{MaMIMO}{massive \gls{mimo}}
\newacronym{PM-MIMO}{PM-MIMO}{post-\gls{mamimo}}
\newacronym{GCoMP}{G-CoMP}{generalized CoMP}
\newacronym{CU}{CU}{central unit}
\newacronym{RU}{RU}{remote unit}
\newacronym{FH}{FH}{fronthaul}
\newacronym{gps}{GPS}{global positioning system}
\newacronym{mse}{MSE}{mean squared error}
\newacronym{LoS}{LoS}{line-of-sight}
\newacronym{TBS}{BS}{base station}
\newacronym{FBS}{FBS}{flying base station}
\newacronym{cpsOFDM}{CPS-OFDM}{circularly pulse-shaped \gls{OFDM}}
\newacronym{pccOFDM}{PCC-OFDM}{polynomial-cancellation-coded \gls{OFDM}}
\newacronym{phy}{PHY}{physical layer}
\newacronym{pls}{PLS}{\gls{phy} security}
\newacronym{PSK}{PSK}{phase shift keying}
\newacronym{gsm}{GSM}{Global System for Mobile Communications}
\newacronym{QAM}{QAM}{quadrature amplitude modulation}
\newacronym{ostbc}{OSTBC}{orthogonal space time block codes}
\newacronym{OOB}{OOB}{out-of-band}
\newacronym{OOBE}{OOBE}{\gls{OOB} emission}
\newacronym{svd}{SVD}{singular value decomposition}
\newacronym{mac}{MAC}{medium access control}
\newacronym{med}{MED}{maximum excess delay}
\newacronym{arq}{ARQ}{automatic repeat request}
\newacronym{OFDM-im}{OFDM-IM}{\gls{OFDM} with index modulation}
\newacronym{fbmc-im}{FBMC-IM}{\gls{fbmc}-index modulation}
\newacronym{c-fbmc-im}{C-FBMC-IM}{Circular convolution \gls{fbmc}-index modulation}
\newacronym{gfdm-im}{GFDM-IM}{\gls{gfdm}-index modulation}
\newacronym{gfdm-sfim}{GFDM-SFIM}{\gls{gfdm} with space and frequency IM}
\newacronym{gfdm-psm}{GFDM-PSM}{\gls{gfdm} with pulse superposition modulation}
\newacronym{OFDM-fsk}{OFDM-FSK}{Universal Filtered OFDM with Filter Shift Keying}
\newacronym{uf-OFDM-im}{UF-OFDM-IM}{Universal Filtered OFDM with Index Modulation}
\newacronym{SM F-OFDM}{SM F-OFDM}{Spatial modulation F-OFDM}
\newacronym{OFDM-DM}{OFDM-DM}{OFDM-differential modulation}
\newacronym{OFDM-SNM}{OFDM-SNM}{OFDM with subcarrier number modulation}
\newacronym{OFDM-PSM}{OFDM-PSM}{OFDM with pulse superposition modulation}
\newacronym{an}{AN}{artificial noise}
\newacronym{FDE}{FDE}{frequency domain equalization}
\newacronym{iot}{IoT}{internet of things}
\newacronym{ACI}{ACI}{adjacent channel interference}
\newacronym{ICI}{ICI}{inter-carrier interference}
\newacronym{ISI}{ISI}{inter-symbol interference}
\newacronym{OFDM-STSK}{OFDM-STSK}{OFDM-aided space-time shift keying}
\newacronym{OFDMa}{OFDMA}{orthogonal frequency division multiple access}
\newacronym{GSFIM-OFDM}{GSFIM-OFDM}{Generalized space frequency index modulation-OFDM}
\newacronym{fmt}{FMT}{filtered multitone}
\newacronym{IM-OFDM-SS}{IM-OFDM-SS}{Index modulated OFDM spread spectrum}
\newacronym{SM}{SM}{Spatial modulation}
\newacronym{SM-OFDM}{SM-OFDM}{Spatial modulation-OFDM}
\newacronym{SM-GFDM}{SM-GFDM}{Spatial modulation-GFDM}
\newacronym{SAP}{SAP}{subcarrier activation pattern}
\newacronym{TD}{TD}{time-domain}
\newacronym{cfo}{CFO}{carrier frequency offset}
\newacronym{to}{TO}{time offset}
\newacronym{pn}{PN}{phase noise}
\newacronym{cfs}{CFS}{carrier frequency shifting}
\newacronym{cir}{CIR}{channel impulse response}
\newacronym{CFR}{CFR}{channel frequency response}
\newacronym{tdd}{TDD}{time division duplex}
\newacronym{ber}{BER}{bit error rate}
\newacronym{ra}{RA}{reconfigurable antenna}
\newacronym{df}{DF}{decay factor}
\newacronym{pdp}{PDP}{power delay profile}
\newacronym{pdb}{PDB}{power difference based}
\newacronym{ds}{DS}{delay spread}
\newacronym{snr}{SNR}{signal-to-noise ratio}
\newacronym{cc}{CC}{\gls{CP} canceling}
\newacronym{mcs}{MCS}{modulation and coding scheme}
\newacronym{RB}{RB}{resource block}
\newacronym{UL}{UL}{uplink}
\newacronym{amc}{AMC}{adaptive modulation \& coding}
\newacronym{lu}{LU}{licensed user}
\newacronym{SDN}{SDN}{software-defined networking}
\newacronym{STA}{STA}{station}
\newacronym{vlc}{VLC}{visible light communication}
\newacronym{led}{LED}{light emitting diodes}
\newacronym{rf}{RF}{radio frequency}
\newacronym{rfid}{RFID}{\gls{rf} identification}
\newacronym{owc}{OWC}{Optical wireless communication}
\newacronym{ir}{IR}{infrared}
\newacronym{vl}{VL}{visible light}
\newacronym{uv}{UV}{ultraviolet}
\newacronym{los}{LOS}{line of sight}
\newacronym{ulaa}{ULAA}{uniform linear antenna array}
\newacronym{fso}{FSO}{free space optical communication}
\newacronym{nlos}{NLOS}{non-line of sight}
\newacronym{uv-lidar}{UV-LIDAR}{ultraviolet light detection and ranging system}
\newacronym{4g}{4G}{fourth generation}
\newacronym{plc}{PLC}{power line communication}
\newacronym{ssl}{SSL}{solid state lighting }
\newacronym{smt}{SMT}{surface mounted technology}
\newacronym{RIS}{RIS}{reconfigurable intelligent surface}
\newacronym{pc}{PC}{phosphorous coated}
\newacronym{apd}{APD}{avalanche photodiode}
\newacronym{pdf}{PDF}{probability density function}
\newacronym{im}{IM}{intensity modulation}
\newacronym{ap}{AP}{access point}
\newacronym{F-AP}{F-AP}{fog access point}
\newacronym{F-UE}{F-UE}{fog \gls{UE}}
\newacronym{NFV}{NFV}{network function virtualization}
\newacronym{nic}{NIC}{network interface card}
\newacronym{oap}{OAP}{optical access point}
\newacronym{wifi}{Wi-Fi}{wireless fidelity}
\newacronym{dsl}{DSL}{digital subscriber line}
\newacronym{pd}{PD}{photodiode}
\newacronym{InGaAs}{InGaAs}{indium gallium arsenide}
\newacronym{ge}{Ge}{germanium}
\newacronym{si}{Si}{silicon}
\newacronym{gbps}{Gbps}{gigabits-per-second}
\newacronym{vlcc}{VLCC}{Visible Light Communication Consortium}
\newacronym{JEITA}{JEITA}{Japan Electronics and Information Technology Industries Association}
\newacronym{ieee}{IEEE}{Institute of Electronics and Electrical Engineers}
\newacronym{ITU}{ITU}{International Telecommunication Union}
\newacronym{dc}{DC}{direct current}
\newacronym{dd}{DD}{direct detection}
\newacronym{pam}{PAM}{pulse amplitude modulation}
\newacronym{ppm}{PPM}{pulse position modulation}
\newacronym{cap}{CAP}{carrier-less amplitude phase}
\newacronym{csk}{CSK}{color shift keying}
\newacronym{lmse}{LMSE}{least mean squared error}
\newacronym{fir}{FIR}{finite impulse response}
\newacronym{ook}{OOK}{on-off keying}
\newacronym{nrz}{NRZ}{non return-to-zero}
\newacronym{fde}{FDE}{frequency domain equalizer}
\newacronym{spam}{SPAM}{superposed PAM}
\newacronym{dppm}{DPPM}{differential PPM}
\newacronym{dappm}{DAPPM}{differential amplitude PPM}
\newacronym{mppm}{MPPM}{multi-pulse PPM}
\newacronym{oppm}{OPPM}{overlapping PPM} 
\newacronym{eppm}{EPPM}{expurgated PPM} 
\newacronym{vppm}{VPPM}{variable PPM}
\newacronym{pwm}{PWM}{pulse width modulation}
\newacronym{u-OFDM}{U-OFDM}{unipolar OFDM}
\newacronym{dco-OFDM}{DCO-OFDM}{DC biased optical OFDM}
\newacronym{aco-OFDM}{ACO-OFDM}{asymmetrically clipped OFDM}
\newacronym{ado-OFDM}{ADO-OFDM}{asymmetrical DC clipped OFDM}
\newacronym{haco-OFDM}{HACO-OFDM}{hybrid asymmetrically clipped optical OFDM}
\newacronym{eu-OFDM}{EU-OFDM}{enhanced Unipolar OFDM}
\newacronym{ifft}{IFFT}{Inverse fast Fourier transform}
\newacronym{laco-OFDM}{LACO-OFDM}{layered ACO-OFDM}
\newacronym{o-noma}{O-NOMA}{optical non orthogonal multiple }
\newacronym{oma}{OMA}{orthogonal multiple access}
\newacronym{tdma}{TDMA}{time division multiple access}
\newacronym{csma/ca}{CSMA/CA}{carrier sense multiple access with collision avoidance}
\newacronym{o-scfdma}{O-SCFDMA}{optical single carrier multiple access}
\newacronym{ocdma}{OCDMA}{optical code division multiple access}
\newacronym{sdma}{SDMA}{space division multiple access}
\newacronym{wdma}{WDMA}{wavelength division multiple access}
\newacronym{hsfo-scfdma}{HFSO-SCFDMA}{Hermitian symmetry free O-SCFDMA}
\newacronym{pdo-noma}{PDO-NOMA}{power domain O-NOMA}
\newacronym{sic}{SIC}{successive interference cancellation}
\newacronym{cdo-noma}{CDO-NOMA}{code domain O-NOMA}
\newacronym{lifi}{Li-Fi}{light fidelity}
\newacronym{emi}{EMI}{electromagnetic interference}
\newacronym{mri}{MRI}{magnetic resonance imaging}
\newacronym{ai}{AI}{artificial intelligence}
\newacronym{v2v}{V2V}{vehicle-to-vehicle}
\newacronym{v2i}{V2I}{vehicle-to-infrastructure}
\newacronym{mrc}{MRC}{maximum ratio combining}
\newacronym{dcn}{DCN}{data center network}
\newacronym{rms}{RMS}{root mean square}
\newacronym{ast}{AST}{adaptive symbol transition}
\newacronym{fcc}{FCC}{Federal Communications Commission}
\newacronym{pts}{PTS}{partial transmit sequences}
\newacronym{psd}{PSD}{power spectral density}
\newacronym{re}{RE}{resource element}
\newacronym{fbmc}{FBMC}{filter bank multicarrier}
\newacronym{SIR}{SIR}{signal-to-interference ratio}
\newacronym{BS}{BS}{base station}
\newacronym{MSE}{MSE-OFDM}{multi-symbol encapsulated OFDM}
\newacronym{UFMC}{UFMC}{universal filtered multicarrier}
\newacronym{f-OFDM}{f-OFDM}{filtered OFDM}
\newacronym{ScS}{$\Delta f$}{subcarrier spacing}
\newacronym{QPSK}{QPSK}{quadrature phase shift keying}
\newacronym{BPSK}{BPSK}{binary phase shift keying}
\newacronym{indxM}{IM}{index modulation} 
\newacronym[plural=RATs,firstplural=radio access technologies (RATs)]{rat}{RAT}{radio access technology}
\newacronym{V2X}{V2X}{vehicle-to-everything}
\newacronym{D2D}{D2D}{device-to-device}
\newacronym{ITS}{ITS}{intelligent transportation system}
\newacronym{UDN}{UDN}{ultra-dense network}
\newacronym{ABS}{ABS}{absolute blank subframe}
\newacronym{TRxP}{TRxP}{transmission reception point}
\newacronym{QoS}{QoS}{quality of service}
\newacronym{QoE}{QoE}{quality of experience}
\newacronym{SWIPT}{SWIPT}{simultaneous wireless information and power transfer}
\newacronym{KPI}{KPI}{key performance indicator}
\newacronym{KQI}{KQI}{key quality indicator}
\newacronym{V-BLAST}{V-BLAST}{Vertical Bell Labs Layered Space-Time}
\newacronym{HPA}{HPA}{high power amplifier}
\newacronym{LLR}{LLR}{log likelihood ratio}
\newacronym{mMTC}{mMTC}{massive machine type communications}
\newacronym{URLLC}{URLLC}{ultra-reliable low latency communications}
\newacronym{OFDM-SPM}{OFDM-SPM}{OFDM with subcarrier power modulation}
\newacronym{OFDM-AIM-FCM}{OFDM-AIM-FCM}{OFDM with adaptive index modulation and fixed constellation modulation}
\newacronym{OFDM-AIM-ACM}{OFDM-AIM-ACM}{OFDM with joint adaptive index modulation and adaptive constellation modulation}
\newacronym{OFDM-VIM-VCM}{OFDM-VIM-VCM}{OFDM with variable index modulation and variable constellation modulation}
\newacronym{TDD}{TDD}{time division duplex}
\newacronym{CQI}{CQI}{channel quality indicator}
\newacronym{RSM}{RSM}{receive SM}
\newacronym{otfs}{OTFS}{orthogonal time frequency and space}
\newacronym{DAC}{DAC}{digital-to-analog converter}
\newacronym{INI}{INI}{inter-numerology interference}
\newacronym{CRS}{CRS}{cell-specific reference signals}
\newacronym{CSI-RS}{CSI-RS}{CSI reference signals}
\newacronym{CSI-IM}{CSI-IM}{CSI interference measurement}

\usepackage{siunitx}
\DeclareSIUnit{\belmilliwatt}{Bm}
\DeclareSIUnit{\dBm}{\deci\belmilliwatt}
\DeclareSIQualifier\isotropic{i}
\DeclareSIQualifier\carrier{c}

\usepackage{array}

\ifCLASSOPTIONcompsoc
 \usepackage[caption=false,font=normalsize,labelfont=sf,textfont=sf]{subfig}
\else
 \usepackage[caption=false,font=footnotesize]{subfig}
\fi

\usepackage[english]{babel}

\usepackage{url}
\usepackage[nameinlink]{cleveref}
\crefname{equation}{}{}
\crefname{section}{Sec.}{Secs.}
\crefname{figure}{Fig.}{Figs.}
\usepackage{balance}
\begin{document}
%
\title{\berkertitle}
%
%
%

\bstctlcite{BSTcontrol}


\author{Murat Karabacak, Berker Peköz, Gökhan~Mumcu\IEEEmembership{,~Member, IEEE} and Hüseyin~Arslan\IEEEmembership{,~Fellow,~IEEE}
\thanks{M. Karabacak was with the Department of Electrical Engineering, University of South Florida, Tampa, FL 33620 USA (e-mail: murat@usf.edu).}%
\thanks{B. Peköz was with the Department of Electrical Engineering, University of South Florida, Tampa, FL 33620 USA. He is now with Wireless Research and Development Division, Qualcomm Technologies Inc., Bridgewater, NJ 08807 USA.}
\thanks{G.~Mumcu and H.~Arslan are with the Department of Electrical Engineering, University of South Florida, Tampa, FL 33620 USA.}}
\maketitle
\begin{abstract}
Advances in computing have resulted in an emerging need for multi-factor authentication using an amalgamation of cryptographic and physical keys. This letter presents a novel authentication approach using a combination of signal and antenna activation sequences, and most importantly, perturbed antenna array geometries. Possible degrees of freedom in perturbing antenna array geometries affected physical properties and their detection are presented. Channel estimation for the plurality of validly authorized arrays is discussed. Accuracy is investigated as a function of \gls{snr} and number of authorized arrays. It is observed that the proposed authentication scheme can provide 1\% false authentication rate at \SI{10}{\decibel} \gls{snr}, while it is achieving less than 1\% missed authentication rates.
\end{abstract}

%
\IEEEpeerreviewmaketitle

\section{Introduction}\label{sec:intro}

\IEEEPARstart{T}{he} emergence of quantum computing has recently shown that currently used conventional encryption techniques can be cracked with ease in the near future \cite{shor_algorithms_1994}. This pushed researchers to finding new horizons that satisfy security requirements through the use of non-cryptographic approaches \cite{bennett_privacy_1999}, such as utilizing the \gls{phy} properties of the system \cite{trappe_challenges_2015} or \gls{ml} techniques  \cite{qadir_adversarial_2019} to infer presence of adversaries and defend accordingly. Quantum password cracking aside, \gls{phy} authentication becomes critical in authenticating simplex broadcasts in which cryptographic approaches cannot be utilized, such as spoofed \gls{gps} signals as in \cite{ranganathan_spree_2016}. In \cite{sharma_survey_2012}, layered security approaches were investigated in detail, and were shown to be redundant and inflexible for future network structures \cite{sharma2011cross}.

Authenticating \glspl{ue} using their \gls{phy} characteristics in developing a \gls{pls} approach have been gaining traction\cite{wang_physical_2016}. The idea of extracting artifacts caused by imperfections in the source \gls{nic} to authenticate devices have been around for more than a decade \cite{brik_wireless_2008}. Channel similarities in addition to the RF fingerprint of the device, of which recent extraction advances is detailed in \cite{xu_device_2016}, are also utilized in the control-layer based authenticator designed in \cite{duan_authentication_2015}, that aims to replace high-latency connections to remotely located authentication servers with local verification among \gls{5g} \gls{hetnet} \glspl{ap}.

Antenna array geometry optimization literature has historically focused on designing "smart" \cite{ertel_overview_1998} or adaptive antenna arrays with improved far- or near-field spatiospectral localization\cite{lebret_antenna_1997}; and is rich in this context. Although \gls{pls} using multiple antennas was also introduced more than a decade ago when signals received from \gls{mimo} transmitters are authenticated using the spatiospectrotemporal correlation of the wireless channel in \cite{xiao_using_2008}. However, due to the randomness of the channel this method can provide limited control on spatiospectrotemporal signatures.  Despite the further studies of PHY security of \gls{mimo} systems in \cite{Hafez2018,pekoz2020}, the literature for \gls{phy} authentication for this systems remains underdeveloped to date. 

Physical layer security aspect of multiple antenna configurations were most recently evaluated to the extent of passive confidentiality and active availability attacks using \gls{mamimo} systems \cite{kapetanovic_physical_2015}. 
Recent developments in \gls{sigint} techniques for \gls{mimo} wireless communications are surveyed in \cite{eldemerdash_signal_2016}. A secure receive spatial modulation scheme that randomizes precoders but not antenna arrays is proposed in \cite{zhang_secure_2019}.

In this work, we propose a novel authentication scheme that combines chaotic antenna array geometries with pseudorandom pilot sequences and antenna array activation sequences. This novel approach combining all three allows unclonable authentication devices, even if the adversaries eavesdrop the message exchange or figure out the unique antenna array geometry by x-ray radiography. The accuracy and scalability of the approach is investigated. It is observed that the proposed authentication scheme can provide 1\% false authentication rate at \SI{10}{\decibel} \gls{snr}, while it is achieving less than 1\% missed authentication rates.

The rest of this paper is organized as follows: Section II provides the adopted system model. Section III introduces the proposed chaotic and pseudorandom designs and briefly presents their effects on the detection metrics. The detection performance results are shown in Section IV. Finally, the paper is concluded in Section V.

Notation: Throughout this paper, vectors are represented using lowercase bold-face letters,
matrices are uppercase bold-face letters, and non-bold letters are used for scalars. The superscripts 
$\herm{\left(\cdot\right)}$
stands for the conjugate-transpose
operation.
$\mathbb{C}$, $\mathbb{Z}$ and $\mathbb{R}$ represent the complex, integer and real number domains, respectively. $\sim\mathcal{CN}\left(\mu,\sigma^2\right)$ corresponds to complex Gaussian distributed random variable with mean $\mu$ and variance $\sigma^2$, and $\mathcal{U} \left( a,b \right)$ corresponds to the uniformly distributed random variable between $a$ and $b$. $\norm{\cdot}{}$ corresponds to the Euclidean norm, $\matr{A}\odot \matr{B}$ and $\matr{A} \oslash \matr{B}$ correspond to the Hadamard multiplication of matrices $\matr{A}$ and $\matr{B}$ and division of matrix $\matr{A}$ to $\matr{B}$, respectively.

The authenticating device and the device being authenticated will hereinafter be referred to as "Seraph" and "Neo", respectively, with subscripts $\cdot_{s}$ and $\cdot_{n}$ used to describe their respective attributes.

\section{System Model}

\begin{figure}
\centering
\includegraphics[width=3.3in]{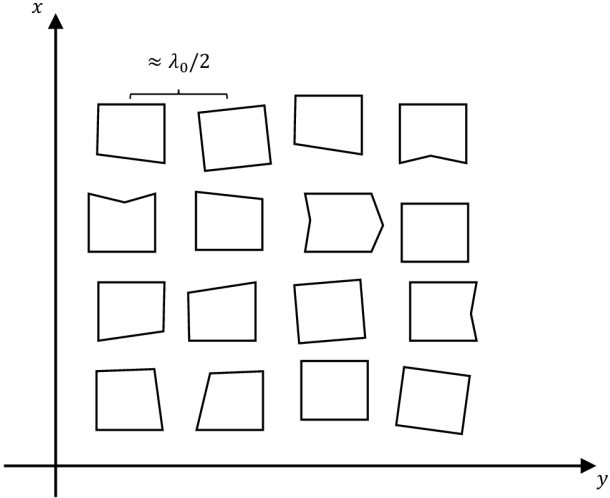}
\caption{An illustration of chaotic antenna array geometry for a $4 \times 4$ antenna array.}
\label{fig:Chaotic}
\end{figure}

The working principle of this system is similar to that of an active \gls{rfid} tag. However, instead of the device-specific variation of binary load state or load impedance as a function of time, it is assumed that upon their first encounter which takes place in a controlled environment, Seraph characterizes and saves the following identifying information about Neo in an allowlist:
\begin{itemize}
    \item The chaotic antenna array geometry equipped by Neo, as exemplified in \cref{fig:Chaotic},
    \item The particular antenna activation sequence used by Neo,
    \item The particular pilot sequence transmitted by Neo.
\end{itemize}
This section contains technical definitions for these information sources. This section and subsequent sections investigate ensuing encounters, during which Seraph authenticates Neo's identity by simultaneously verifying all abovementioned attributes of Neo real-time in uncontrolled environments. The analysis further assumes that, as is the case with \gls{rfid} tags, Seraph and Neo are synchronized; and the wireless propagation channel between each antenna of Seraph and Neo is representable in the form of a single tap over the utilized bandwidth without loss of generality, is also time-invariant throughout the transmission interval, and is known by Seraph through readily available techniques.

Neo is equipped with $M_n=H_n\times V_n\in \mathbb{N}$ antennae wherein $H_n\in \mathbb{N}$ and $V_n\in \mathbb{N}$ correspond to the number of antennae on the horizontal and vertical edges of Neo's 2D antenna array. Neo's 2D antenna array starts off as a standard $\lambda_0 /2$ spaced \gls{ulaa}, where $\lambda_0$ is the free space wavelength at the center carrier frequency. Each antenna element starts off as square patch of edge length $\lambda_g /2$, where $\lambda_g<\lambda_0$ is the guided wavelength at the center carrier frequency, and each vertex of each antenna element is translated from its original location as $
 p_{m,\alpha}=\bar{p}_{m,\alpha}+u_{x,m,\alpha}\hat{i}+u_{y,m,\alpha}\hat{j}$,
where $p_{m,\alpha}$ is the final coordinate of the $\alpha\in\mathbb{Z}^{+}_{\leq 4}$th the vertex of the $m\in\mathbb{Z}^{+}_{\leq M_n}$th antenna element, $\bar{p}_{m,\alpha}$ is the original coordinate thereof, $u_{x,m,\alpha}$ and $u_{y,m,\alpha}$ are both $\sim \mathcal{U}\left(-\lambda_g /4,\tfrac{\lambda_0-\lambda_g}{4}\right)$ and denote the horizontal and vertical displacement of the aforementioned vertex from its original location, respectively, and $\hat{i}$ and $\hat{j}$ are the horizontal and vertical unit length vectors, respectively. Furthermore, the joint \gls{pdf} for any two displacement satisfies
$ f_U\left(u_{\beta_0,m_0,\alpha_0},u_{\beta_1,m_1,\alpha_1}\right)=f_U\left(u_{\beta_0,m,\alpha_0}\right)f_U\left(u_{\beta_1,m,\alpha_1}\right)\forall \beta_{0,1}\in\{x,y\};m_{0,1}\in\mathbb{Z}^{+}_{\leq M_n};\alpha_{0,1}\in\mathbb{Z}^{+}_{\leq 4}$.
 Note that by independently displacing all vertices in two dimensions, each antenna element is translated, rotated, scaled or skewed chaotically from the \gls{ulaa} design as illustrated in Fig. \ref{fig:Chaotic}. As a result, it is assumed that complex noise is introduced to Neo's spatial signature in the transmit direction of $\Omega$ \cite{tse_mimoi_2005} as 
 \begin{IEEEeqnarray}{C}
      \matr{h}_n \left( \Omega \right)=\frac{\matr{h}\left( \Omega \right)+\sigma_{\matr{h}}\Tilde{\matr{h}}\odot\matr{e}_{\mathrm{t}}\left( \Omega \right)}{\sqrt{2}},
 \end{IEEEeqnarray}
wherein $\matr{h}\left( \Omega \right)\in\mathbb{C}^{M_n\times 1}$ is the spatial signature of Neo's nonmodified \gls{ulaa} in the transmit direction of $\Omega$ as described in \cite[Eq. (7.24)]{tse_mimoi_2005} of which construction is not recited here due to space constraints, $\sigma_{\matr{h}}\in \mathbb{R}^{+}$ is the standard deviation thereof and corresponds to the positive square root of the channel gain, $\matr{e}_{\mathrm{t}}\left( \Omega \right)\in\mathbb{C}^{M_n\times 1}$ is Neo's nonmodified \gls{ulaa}s unit spatial signature in the transmit direction of $\Omega$ as described in \cite[Eq. (7.25)]{tse_mimoi_2005}, and $\Tilde{\matr{h}}\in\mathbb{C}^{M_n\times 1}$ is the introduced chaotic noise of which each element is $\sim\mathcal{CN}\left(0,1\right)$ independent from others. Accordingly, we will refer to Neo's final \gls{ulaa}s unit spatial signature in the transmit direction of $\Omega$ as 
\begin{IEEEeqnarray}{C}
 \matr{e}_{\mathrm{n}}\left( \Omega \right) = \frac{\sqrt{2}\matr{h}_n \left( \Omega \right)-\matr{h} \left( \Omega \right)}{\sigma_{\matr{h}}}\oslash \Tilde{\matr{h}}.\label{eq:neounitspasig}
\end{IEEEeqnarray}

Each antenna is connected to an independent RF chain that is capable of carrying a complex (IQ modulated) sinusoid pulse uncorrelated to those of other antennae. The reciprocal of the duration of each pulse is analogous to widely known "baud rate" and is assumed constant, at least for Neo, to ease practical aspects concerning transceiver implementation. Neo may also utilize nonsinusoidal wavelets, or, further utilize plurality of wavelets wherein each signal element utilizes a different wavelet for further scalability and security, but these are beyond the scope of this art and will be considered in future works to maintain the work in hand concise. Neo transmits the pilot sequence over $T_n\in\mathbb{N}$ baud intervals, and the pilot symbol modulating the sinusoid transmitted from the $m\in \mathbb{Z}^{+}_{\leq M_n}$th antenna during the $t\in \mathbb{Z}^{+}_{\leq T_n}$th baud interval is given in the $m$th row and $t$th column of the pilot matrix $\matr{X}_n\in\mathbb{C}^{M_n\times T}$ and denoted by $\matr{X}_{n}\left(m,t\right)$, wherein 
$\angle \matr{X}_{n}\left(m,t\right) \sim \mathcal{U} \left( -\pi,\pi \right) $ and 
\begin{IEEEeqnarray}{C}
 \left| \matr{X}_{n}\left(m,t\right) \right| =\begin{cases} \sim \sqrt{\mathcal{U}\left( 0,1 \right)} & \text{, } \nu_{n,m,t} \geq \nu_{n} \\ 0 & \text{, o.w.} \end{cases},
\end{IEEEeqnarray}
where $\nu_{n,m,t} \sim \mathcal{U}\left( 0,1 \right) $ is a random variable that determines whether Neo's $m$th antenna during the $t$th baud interval and $\nu_n$ is Neo's activation threshold that determines the antenna activation probability; furthermore $ f_{P}\left( \nu_{n,m_0,t_0} , \nu_{n,m_1,t_1} \right) = f_{P}\left( \nu_{n,m_0,t_0}  \right) f_{P}\left(  \nu_{n,m_1,t_1} \right) \forall m_0,m_1 \in \mathbb{Z}^{+}_{\leq M_n}; t_0,t_1 \in \mathbb{Z}^{+}_{\leq T_n} $. A zero entry in $\matr{X}_n$ implies that no transmission occurs from that antenna during that baud interval. Therefore, $\matr{X}_n$ describes the particular antenna activation sequence used by Neo in its columns, and the particular pilot sequence transmitted by Neo in its elements.

Seraph is equipped with $N_s\in\mathbb{N}$ antennae that is formed in a nonmodified 2D \gls{ulaa}, and has the default spatial signature thereof. Accordingly, the channel matrix $\matr{H}_n\in\mathbb{C}^{N_s \times M_n}$ is composed as done in \cite[Eq. (7.56)]{tse_mimoi_2005}, with the difference being the unit spatial signature in the transmit direction of $\Omega$ term denoted by $\matr{e}_{\mathrm{t}}\left( \Omega \right)$ is replaced with $\matr{e}_{\mathrm{n}}\left( \Omega \right)$ derived in \eqref{eq:neounitspasig}. The signal received at Seraph's $n<N_s$th antenna at the end of the $t<T_n$th baud interval is given on the $n$th row and $t$th column of $\matr{y}_s\in \mathbb{C}^{N_s \times T_n}$, where
\begin{IEEEeqnarray}{C}
 \matr{y}_s=\matr{H}_n \matr{X}_n + \matr{w},
\end{IEEEeqnarray}
where $\matr{w}\in \mathbb{C}^{N_s \times T_n}$ is the \gls{awgn} matrix comprising independent elements identically distributed with $\sim\mathcal{CN}\left( 0,\left(\sigma_{\matr{h}}/\gamma_n\right)^{2} \right)$ wherein $\gamma_n$ is Neo's \gls{snr}.
\section{Proposed Receiver}

Since Seraph relies on random deviations of spatial signature, a detection algorithm for Neo's spatial signature deviation can be implemented to decide if the received signal is coming from Neo or not. The detection algorithm can be derived by correlating Neo's expected received signal over $\matr{y}_s$. The correlation is calculated using
\begin{IEEEeqnarray}{C}
 \rho =\mathrm{tr} \left( \herm{\matr{X}_n}\herm{\matr{H}_n}\matr{y}_s \right).
\end{IEEEeqnarray}

Noise or signal emitted by intruders may also cause high correlation and can cause false positives if solely this measurement is considered. To distinguish Neo's signal both from noise and signals emitted by intruders, its strength must be compared to that of noise. 
The noise variance, $\hat{\sigma}_n^{2}\in\mathbb{R}$, is estimated by similarly correlating $\matr{y}_s$ with any signature orthogonal to that of Neo and all other possible authorized users. The detection metric $\beta\in\mathbb{R}$ is then given by
\begin{IEEEeqnarray}{C}
 \beta = \frac{\rho}{\hat{\sigma}_n^{2}}.
\end{IEEEeqnarray}
The detection metric is then compared to a threshold value ($\psi$). To minimize the error, one threshold can be selected as half of the distance between two states as 
\begin{IEEEeqnarray}{C}
\psi_e =\frac{ \mathrm{tr} \left( \herm{\matr{X}_n}\herm{\matr{H}_n}\matr{H}_n\matr{X}_n \right)}{2\hat{\sigma}_n^{2}}.
\end{IEEEeqnarray}
However, $\psi_e$ is not a good threshold for low \gls{snr} scenarios, which results in high \gls{fa} rates. To prevent that, a threshold ($\psi_{FA}$) can be precalculated to fix the \gls{fa} probability to a designed value. \Gls{fa} probability can be represented as
\begin{IEEEeqnarray}{C} \label{probFA}
\Pr \left( \mathrm{FA} \right) = \Pr \left(\beta > \psi |\matr{y}_s=\matr{w} \right)
\end{IEEEeqnarray}
which is $\sim\mathcal{N} \left( 0,\mathrm{tr} \left( \herm{\matr{X}_n}\herm{\matr{H}_n}\matr{H}_n\matr{X}_n \right) \right)$. The final threshold is found as the combination of both thresholds as
\begin{IEEEeqnarray}{C} \label{probFA}
\psi = \mathrm{max}(\psi_e,\psi_{FA})
\end{IEEEeqnarray}
to improve the performance of the system. Performance analysis for a variety of false alarm thresholds is presented in \cref{sec:perf}.

\section{Performance Analysis}\label{sec:perf}

To evaluate the proposed authentication method, link level simulations have been performed under highly scattering Rayleigh channel. Seraph is assumed to have 512 antennas at all times, while Neo may have different number of active antennas depending on $\eta_{n}$.

\begin{figure}
\centering
\includegraphics[width=3.3in]{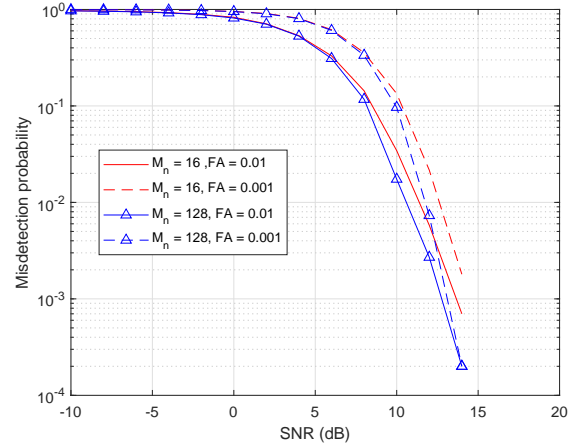}
\caption{Missed detection rate for various $M_n$ and \gls{fa} targets.}
\label{fig:miss}
\end{figure}

In \cref{fig:miss}, missed detection rate of Neo's signature is presented against \gls{snr}. The proposed method fails to authenticate Neo with less than 1\% probability at most at \SI{13}{\decibel} \gls{snr} if $M_n=16$ antennas are activated while $\Pr \left( \mathrm{FA} \right) = 0.001$. It is also seen that 1\% misdetection probability can be achieved when \SI{8}{\decibel} \gls{snr} for both $M_n=16$ and $M_n=128$ active antennas with relaxed FA requirement of $\Pr \left( \mathrm{FA} \right) = 0.01$. Lower rates are possible as the number of active antennas are increased or \gls{fa} probability requirement is relaxed. 

\begin{figure}
\centering
\includegraphics[width=3.3in]{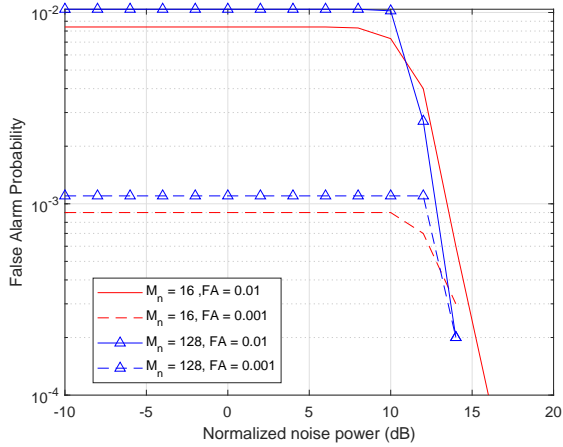}
\caption{\gls{fa} rate when receiving only noise, for various $M_n$ and \gls{fa} targets.}
\label{fig:false1}
\end{figure}

\cref{fig:false1} shows \gls{fa} rates of Neo's signature when Seraph is only receiving noise. 
As it is seen from the figure, the desired FA rates of $\Pr \left( \mathrm{FA} \right) = 0.001$ and $\Pr \left( \mathrm{FA} \right) = 0.01$ are closely achieved for \gls{snr} values up to \SIlist{12;10}{\decibel}, respectively. After that, \gls{snr} values the secondary threshold of $\psi_e$ becomes effective and improves the \gls{fa} performance for Neo's signature detection algorithm. The authors note that, due to the designed threshold calculation based on the desired false detection rate, both results are expected to be as close as possible to $10^{-2}$ or $10^{-3}$ depending on the threshold. Since higher number of antennas provide better diversity against fading and noise, increasing the number of antennas provides more consistent behaviour that yield results closer to expected. Lower number of antennas may result in reduced false alarm rates due to the way the detector is structured, but increasing number of antennas results in more consistent overall detection performance also considering the misdetection rates shown in \cref{fig:miss}.

\begin{figure}
\centering
\includegraphics[width=3.3in]{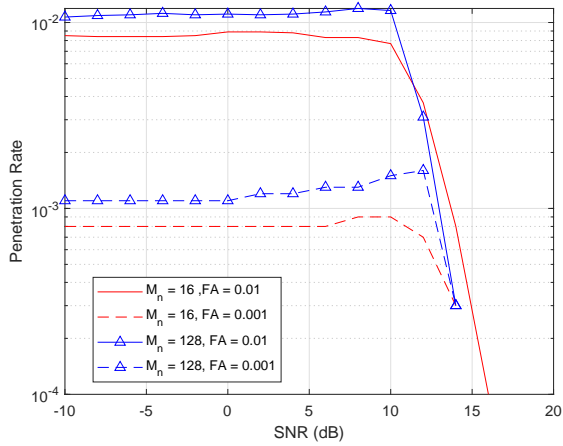}
\caption{\gls{fa} rate when receiving a random signature from a random transmitter (intruder) for various $M_n$ and \gls{fa} targets and intruder's \gls{snr}.}
\label{fig:false2}
\end{figure}

The penetration rate of a random signature intruder is presented in \cref{fig:false2}. In this scenario, Seraph receives a random signal from a transmitter which has a random signature that is different than Neo's signature. As seen in \cref{fig:false2}, penetration test performance shows similar behaviour to the noise-only scenario presented in \cref{fig:false1}. Due to increased signal power and \gls{fa} rates being fixed for noise only scenario, \gls{fa} rates slightly increases at around \SI{10}{\decibel} \gls{snr}. Then similar to previous case, after \gls{snr} values of \SI{12}{\decibel} and \SI{10}{\decibel} for $\Pr \left( \mathrm{FA} \right) = 0.001$ and $\Pr \left( \mathrm{FA} \right) = 0.01$, respectively, the secondary threshold of $\psi_e$ becomes effective and improves the \gls{fa} performance.

\balance
\section{Concluding Remarks}

A novel authentication approach combining chaotic antenna array geometries with signal and antenna activation sequences has been presented. Possible degrees of freedom in perturbing antenna array geometries, affected physical properties and their detection are presented. While enforcing false alarm rate to be less than 1\%, the proposed authentication method is able to provide less than 1\% missed detection rates above It is observed that the proposed authentication scheme can provide 1\% false authentication rate at \SI{8}{\decibel} \gls{snr}. Practical approached to randomized chaotic antenna array manufacturing and statistical signature distribution of the manufactured arrays can be investigated as a future study.

\ifCLASSOPTIONcaptionsoff
  \newpage
\fi



%

\bibliographystyle{jabbrv_IEEEtran}
\bibliography{IEEEabrv,berker,3gpp_38-series}

\end{document}